\documentclass{article}
\usepackage[utf8]{inputenc}
\usepackage[letterpaper, margin=1in]{geometry}
\usepackage[T1]{fontenc}
\usepackage{microtype}
\usepackage{graphicx}
\usepackage{caption}
\usepackage{subcaption}
\usepackage{booktabs} 
\usepackage{mathtools}
\usepackage{times}
\usepackage{helvet}
\usepackage{courier}
\usepackage[hyphens]{url}
\usepackage{xcolor}
\usepackage{color, colortbl}

\usepackage{bbm}

\usepackage{epsfig}
\usepackage{pst-grad} 
\usepackage{pst-plot} 
\usepackage[space]{grffile} 
\usepackage{etoolbox} 
\makeatletter 
\patchcmd\Gread@eps{\@inputcheck#1 }{\@inputcheck"#1"\relax}{}{}
\makeatother

\usepackage{amsfonts,amsthm,amssymb}
\usepackage{amsmath}

\usepackage{nicefrac}
\usepackage{algorithm}
\usepackage[noend]{algpseudocode}
\usepackage{enumerate}
\usepackage{CJK}

\usepackage{mdframed}

\usepackage{multirow}
\usepackage{longtable}

\newtheorem{lemma}{Lemma}[section]
\newtheorem{theorem}[lemma]{Theorem}

\newtheorem{corollary}[lemma]{Corollary}

\allowdisplaybreaks

        {\hspace*{\fill}$\Box$\par}

\title{Instance Optimal Join Size Estimation} 
\author{Mahmoud Abo-Khamis \\ RelationalAI\\
mahmoud.abokhamis@relational.ai\and Sungjin Im~\thanks{Supported in part by NSF grants CCF-1409130, CCF-1617653, and CCF-1844939.}\\ University of Callifornia, Merced\\sim3@ucmerced.edu \and Benjamin Moseley~\thanks{Supported in part by a Google Research Award, an Infor Research Award, a Carnegie Bosch Junior Faculty Chair and NSF grants CCF-1824303,  CCF-1845146, CCF-1733873 and CMMI-1938909.} \\Carnegie Mellon University \\ moseleyb@andrew.cmu.edu \and Kirk Pruhs~\thanks{Supported in part  by NSF grants  CCF-1535755, CCF-1907673,  CCF-2036077 and an IBM Faculty Award.} \\ University of Pittsburgh \\ kirk@cs.pitt.edu \and Alireza Samadian \\ University of Pittsburgh \\ samadian@cs.pitt.edu}

\date{}
\usepackage{graphicx}

\begin{document}
\maketitle
\begin{abstract} 
 We consider the problem of efficiently estimating the size of the inner join of a collection  of preprocessed relational tables from the perspective of instance optimality analysis. 
The run time of instance optimal algorithms is  comparable to the minimum time needed to verify the correctness of a solution.
Previously instance optimal  algorithms were only known when the size of the join was small (as one component of their run time that was linear in the join size). 
We give an instance optimal algorithm for estimating the join size for all instances, including when the join size is large, by removing the dependency on the join size. As a byproduct we show how to sample rows from the join uniformly at random 
in a comparable amount of time. 
\end{abstract}

\section{Introduction}

\subsection{The Problem Statement}

We consider the problem of  efficiently estimating the size, or equivalently the number
of rows, in an  inner join
$$J = T_1 \Join T_2 \Join \ldots \Join T_t$$
given as input the collection $\tau = \{T_1, \ldots T_t \}$ of relational tables, and some associated data structures
$D_1, \ldots, D_t$ that were obtained by independently 
preprocessing the tables. 
We also consider the  related problem of 
uniformly sampling a row from $J$.

One application of estimating the join size is in the context of determining a physical query plan for
a particular logical query.~\cite[Chapter 15]{Ullman}
As example, let us consider the logical query $T_1 \Join T_2 \Join T_3$. 
Assume that one knew that $T_1 \Join T_2$ was large, but 
$T_2 \Join T_3$ and $T_1 \Join T_2 \Join T_3$ were small. 
Then first joining 
$T_2$ with $T_3$ and then joining the result with $T_1$,  would likely be a more efficient 
physical query plan than first joining $T_1$ and $T_2$ as it would avoid (unnecessarily) computing
a large intermediate result. 
If the joined table $J$ is large, one common method for efficiently solving some computational
problem on a joined table $J$ is to solve the problem on a uniformly
sampled collection of points from $J$. For example, \cite{unconditionalCoreset} shows
that uniformly sampling can be used to efficiently solve a wide range of common 
regularized loss
minimization problems with arbitrary accuracy. 

It is NP-hard to determine whether or not the the joined table $J$ is empty (see for example \cite{grohe2006structure,marx2013tractable}). Thus it is NP-hard to
approximate the join size within any reasonable factor. 
Due to this hardness result, this paper addresses join size  estimation from the perspective of
instance optimality analysis. 

\subsection{Instance Optimality Analysis}

In instance optimality
analysis one ideally seeks an 
algorithm $A$ that on every instance $\cal I$ is
almost as fast as the fastest correct algorithm $A$ is
on instance $\cal I$. That is,
$A({\mathcal I}) \le \min_{A' \in {\mathcal C}} A'({\mathcal{I}})$,
where $A({\mathcal I})$ is the time for algorithm $A$ on
input $\mathcal I$, $A'({\mathcal I})$ is the time for algorithm $A'$ on
input $\mathcal I$, and $\mathcal C$ is the collection 
of all correct algorithms. Note that in order for
an algorithm to be correct it must be correct on
all inputs, not just input $\mathcal I$. 

However this ideal is 
overly ambitious and impractical for most problems
for several reasons. One of these reasons is that
characterizing  the fastest time that an 
algorithm can solve a particular problem or
problem instance is generally
well beyond our current understanding (the P=NP problem
is essentially the problem of characterizing the
fastest time that instances of NP-complete
problems can be solved). 
A standard workaround 
is to first observe that the nondeterministic time complexity lower
bounds the deterministic time complexity.
That is the fastest time that an algorithm
can solve an 
instance 
is lower bounded by the fastest time that an algorithm
can verify a proof or certificate that
a candidate solution is correct for that instance. 
Again allowing arbitrary certificates is overly
ambitious and impractical. So generally for each 
problem one picks a particular type $\mathcal C$ 
of certificate 
that seems natural for the particular problem under
consideration. Then one might seek an algorithm $A$ where
the time of $A$ is at most the minimum certificate size. 
That is, 
$A({\mathcal I}) \le  \min_{C \in {\mathcal C}({\mathcal I})} |C({\mathcal I})|$,
where here ${\mathcal C}({\mathcal I})$ is the collection of
certificates of type $\mathcal C$ that proves
that a particular answer is correct for the instance
$\mathcal I$, and $|C|$ denotes the size of $C$.
Such an algorithm $A$ would then be instance optimal 
among all algorithms that produce a certificate of type $\mathcal C$. For example, \cite{AfshaniBC17} gives an algorithm
for finding Pareto frontier of points in the plane that is
instance optimal among comparison based sorting algorithms.

One often needs to relax this a bit more via approximation.
So we seek algorithms $A$ where
 $$A({\mathcal I}) = \tilde O(  \min_{C \in {\mathcal C}({\mathcal I})} F(|C({\mathcal I})|))$$
 where the $\tilde O$ hides poly-log factors of the input size,
 and $F$ is some function that is as close to linear as possible.
 That is, ignoring poly-log factors, we want
 the time to be a function
 of the certificate size, and we want this function to be as slowly growing
 as possible. 
 
 Instance optimality analysis is often viewed as falling under the
rubric of ``beyond worst-case'' analysis techniques. A survey
of instance optimality analysis and other ``beyond worst-case''
analysis techniques can be found at \cite{Roughgarden19}.

\subsection{Previous Instance Optimality Results for
 Join Size}

Here we consider a type of 
certificate that has been used in previous papers
on join size computation using instance optimality analysis~\cite{khamis2016joins,alway2019domain,abs-1909-12102},
namely a collection of gap boxes. 
To apply this certificate one assumes that a priori 
it is the case that all table entries have been mapped
to integers in the range $[1, n]$. 
Thus if there are a total of $d$ attributes in the 
tables in $\tau$ then $J$ can be viewed as a collection
of $z$ axis-parallel unit hypercubes in $[0,n]^d$.  To accomplish
this, one can view each 
row $r$ of $J$ as a point in ${\mathbb N}^d$ in the natural way, and then associate
$r$ with the axis-parallel unit hypercube $h$ that contains $r$,
and where $r$ is the furthest point in $h$ from the origin. 
We define a box to be 
an axis parallel hyperrectangle that is the Cartesian
product of intervals. Then a gap box $b$ is a box
that does not intersect any unit hypercube in $J$.
A collection $B$ of gap boxes, whose union has
volume $V$, can be viewed
as a certificate that the join size $z$ is at
most $n^d - V$. Let $C$ be the minimum cardinality collection
of gap boxes that cover $[0, n]^d - J$. 

\begin{figure}
    \centering
    \begin{subfigure}[b]{0.3\textwidth}
        \centering
        \includegraphics[scale=0.1]{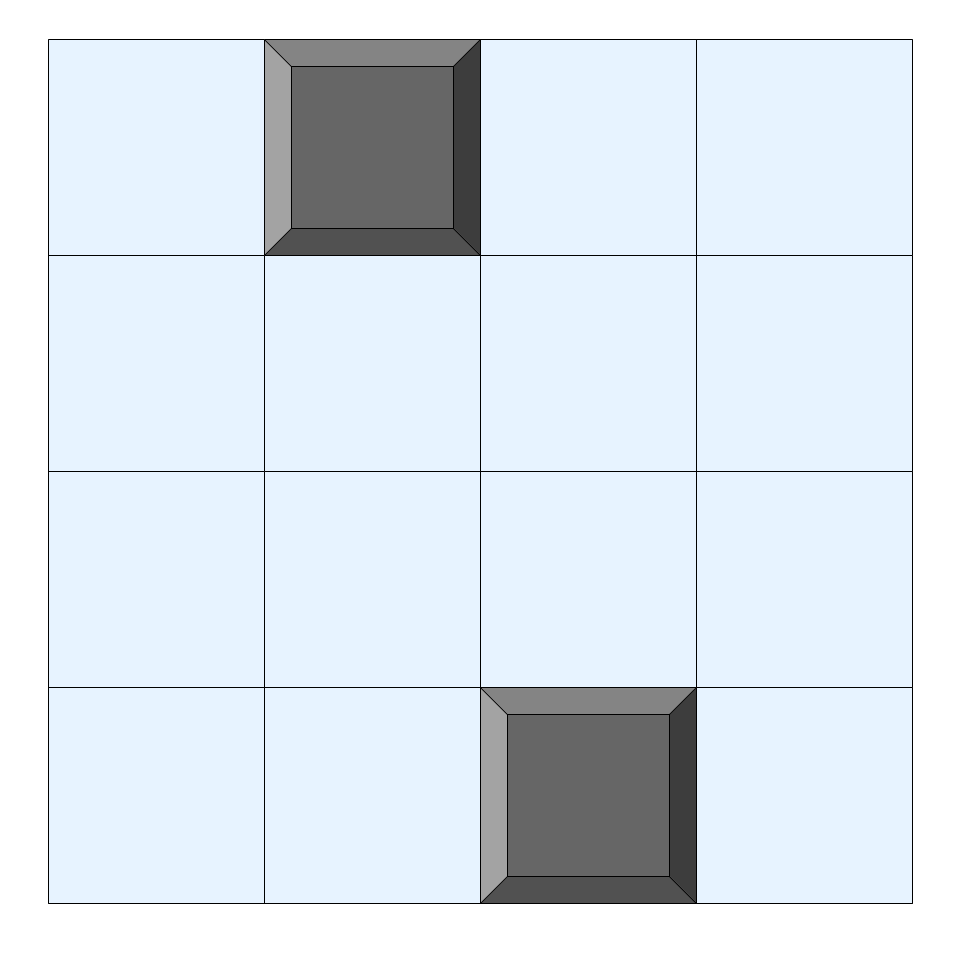}
        \caption{}
        \label{fig:points}
    \end{subfigure}
    \begin{subfigure}[b]{0.3\textwidth}
        \centering
        \includegraphics[scale=0.1]{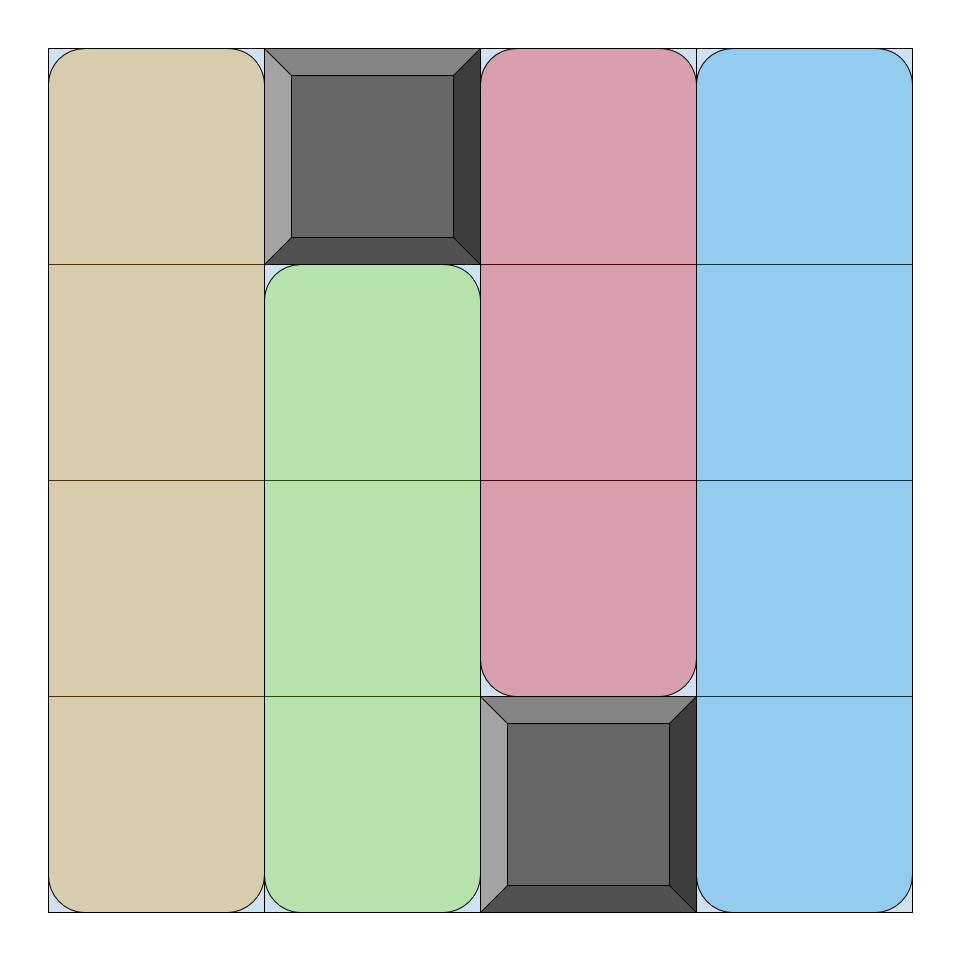}
        \caption{}
        \label{fig:boxes}
    \end{subfigure}
    \caption{Two points in two dimensional space and the minimum number of gap boxes needed to cover the empty space.}
    \label{fig:gapexample}
\end{figure}

As an example of these definitions, consider the instance in Figure \ref{fig:gapexample} where $d=2$.
 Figure \ref{fig:points} shows the two unit hypercubes associated with the points
$(2, 4)$ and $(3, 1)$. Figure \ref{fig:boxes} shows
a minimum cardinality collection of gap boxes that 
establishes that volume of these points is at most two.
So $C=4$ for this instance.

\cite{khamis2016joins} gives an algorithm Tetris for the join size problem.
The algorithm Tetris and its analysis was extended in \cite{alway2019domain,abs-1909-12102}.
\cite{alway2019domain,abs-1909-12102} shows how to preprocess each table to construct
a collection of $O(d n \log^d n)$  gap boxes, which are the maximal
dyadic gap boxes. 
A {\em dyadic box} is a box that its width in each dimension is a power of $2$ and its projection into  each dimension is an integer multiple of its width in that dimension.
\cite{alway2019domain,abs-1909-12102} shows that any certificate of
gap boxes can be replaced by a certificate that is a subset of maximal dyadic gap boxes without a substantial increase in the size of the certificate. The running time of the 
preprocessing is $\tilde{O}(n)$. \cite{khamis2016joins} explains how to store
dyadic gap boxes in what is essentially a trie using space $O(d \log n)$
per box. 
Using this data structure, Tetris  computes the exact join size
$z$ in time $\tilde{O}(|C|^{d/2} + z)$ by
constructing a collection $B$ of $\tilde O(C)$ gap boxes.
Conceptually in each iteration Tetris finds a point $p$ that is not in
the joined table $J$, and not covered by any box in $B$, and then
adds to $B$ all the dyadic boxes that contain $p$  in any of
the data structures associated with any of the tables. 
The costly operation is finding $p$, which conceptually is achieved
by exhaustive enumeration. 

\subsection{Our Results}

In  many applications of the join size computation/estimation, 
such as the application of computing a physical query plan,
the linear dependence on the join size $z$ in the running time of Tetris is not desirable.
In these settings one wants to be able to estimate quickly even if the join is large. 
Our main result is an algorithm Welltris~\footnote{Welltris the game is a variation of the 
original game Tetris~\cite{Welltris}} that is a variation of 
Tetris, that estimates the join size without any dependence in the running
time on the join size $z$. 
So we assume that the input to the join size
problem additionally
contains two parameters $\epsilon$ and $\delta$.
Welltris then computes an estimate $\tilde z$ to the number of rows $z$ in the joined table $J$
that is accurate to within a factor of $1+\epsilon$ with probability at least $1-\delta$. That is
$\Pr\left[ \frac{z}{1+\epsilon} \le \tilde z \le (1+\epsilon) z\right] \ge 1- \delta$.

The preprocessing of the tables in Welltris is identical to that of Tetris used in
\cite{alway2019domain,abs-1909-12102}.
But for completeness we cover this in section \ref{sect:preprocessing}. 
When computing the join size, Welltriss, like Tetris, maintains a collection $B$
of $\tilde O( C)$ dyadic boxes, and in each iteration it finds a point $p$ not in
the joined table and not covered by any box in $B$. But Weltriss does not
use exhaustive search to find $p$. Instead Welltris \emph{samples} points uniformly
at random from points not covered by the boxes in $B$. The intuition behind this is that the uniform sampling will likely lead to choosing choosing boxes that contain a large number of points uncovered by any boxes in $B$. To accomplish this it 
modifies Chan's algorithm~\cite{chan2013klee} for Klee's measure problem, which is
the problem of computing the volume of the union of a collection of boxes. 
The running time for Chan's algorithm, which is $\tilde O(  C^{d/2})$.  
We show that the running time of 
Welltris is $\tilde{O}(C^{d/2+1} )$~\footnote{This assumes that $\epsilon$, $\delta$ and $d$ are treated as constants.}, or more precisely, 
$$O\left(    C^{d/2+1} \cdot \frac{1}{\epsilon} \cdot   \log(1/\delta) \cdot 2^{d^2/2+d} \cdot \log^{d^2+d}(n) \cdot \log(m) \right)$$
So Welltris can estimate the join size quickly for instances in which there
is a modestly-sized certificate bounding the join size, and in which the number of dimensions is small,
even if the join size is large.

Welltris can be modified to sample $q$ points uniformly from the joined table $J$ in time  $\tilde{O}( C^{d/2+1} + q C)$, or more precisely, 
$$O\left( C^{d/2+1}   \cdot \log(q) \cdot 2^{d^2/2+d} \cdot \log^{d^2/2+d}(n) \cdot \log^{d^2/2}(m)  + q \cdot C \cdot d^4 2^d \cdot \log^{2d+1}(n) \right)$$

It is $W[1]$ hard to 
decide if a collection of boxes covers the whole space,
that is to decide whether $z=0$ or whether $z > 0$~\cite{CHAN2010243}. 
Thus a running time that did 
not have exponential dependence on $d$ would imply $W[1]=FPT$,
which is considered unlikely.
For this reason, believe Welltris   has roughly
optimal efficiency for join size estimation with this type of certificate.

\section{Formal Definitions}
In this section we state the formal definitions of the terms used in the rest of the paper. 
A \emph{table/relation} $T_i$ with $d_i$ columns/dimensions as a collection of points in a $d_i$ dimensional space such that each dimension has an associated
positive integer that we refer to as the attribute of the column. 
No attribute can be associated with more than one dimension in the table. 
Different tables may share  attributes. 
The \emph{join} of $T_1,\dots, T_t$ is a table $J$ such that the set of attributes of $J$ is the union of the attributes of $T_1,\dots, T_t$ and a point $x$ is in $J$ if and only if for all $T_i$ the projection of $x$ onto the subspace formed by the columns of $T_i$ exists in $T_i$.
We use $t$ to denote the number of tables in our join query, $d$ to denote the total number of columns, $m$ to be the maximum number of rows in any of the input tables, and $n$ to be the size of the domain of each dimension (the maximum number of unique values in that dimension). Note that if we assume that the tables involved in the join are the only tables in the database, then $n \leq t m$.

Given a set of points in $d$ dimensional space, a {\em gap box} is a $d$ dimensional box that does not contain any of the points. We denote the set of all gap boxes by $B$. 
$b \in B$ is a {\em maximal gap box} if and only if there is no other box in $B$ that covers $b$ entirely and is not equal to $b$. We denote the set of all maximal boxes in $B$ by $B_m$.
Given a set of boxes $B'$, we call $C(B')$ a certificate for $B'$ if and only if it is a set of minimum number of boxes needed to cover the union of $B'$.
It is easy to observe that if $B_m$ is the set of maximal boxes in $B$ then $|C(B)| = |C(B_m)|$.

A {\em dyadic gap box} is a gap box that its size in each dimension is a power of $2$ and its location in each dimension is a multiple of its size in that dimension. We show the set of maximal dyadic gap boxes by $B_d$.
A dyadic box can be represented by a $d$ dimensional vector and in each dimension it has the binary prefix of a set of points; then such a dyadic box will cover all the points that this box is a prefix of in all the dimensions. Note that the binary prefix can be empty in a dimension (denoted by $\lambda$) and in that case that dyadic box spans that dimension entirely. Note that each point is also a dyadic box of size 1.
For $d=3$ and $n=8$, a dyadic box can be $(\lambda,1,101)$ which means its size in the first dimension is $8$, in the second one $4$, and in the third dimension is $1$. In this case, $(111,100,101)$ is a point covered by this dyadic box and $(111,000,101)$ is an uncovered point.

\begin{lemma}[\cite{khamis2016joins}]
\label{lemma:dyadicbox_containing_dyadicbox}
Each dyadic box can be fully contained in at most $\log^d(n)$ other dyadic boxes.
\end{lemma}
\begin{proof}
Based on definition of dyadic box, a dyadic box $A = (a_1,\dots, a_d)$ contains all the points $x = (x_1,\dots,x_d)$ such that $a_i$ is a prefix of $x_i$ for all $i$. Therefore, a dyadic box $A = (a_1,\dots, a_d)$ contains a dyadic box $B=(b_1,\dots, b_d)$ if and only if there $a_i$ is a prefix of $b_i$ for all $i$. Since each dimension has at most $log(n)$ prefix, the lemma follows.
\end{proof}

\begin{lemma}[\cite{alway2019domain,abs-1909-12102,khamis2016joins}]
The maximum number of dyadic boxes needed to cover the area of a gap box is at most $2^d\log^d(n)$.
\end{lemma}



\section{Preprocessing}
\label{sect:preprocessing}

\subsection{Gap box construction}
The preprocessing step for our algorithm is constructing the dyadic gap boxes for all the input tables. The algorithm introduced in \cite{alway2019domain} produces all maximal dyadic boxes of the input tables. It is also shown that any certificate can be replaced by a certificate that is a subset of maximal gap boxes without a substantial increase in the size of the certificate. For completeness, we explain the algorithm in \cite{alway2019domain} and the relevant theorems.

The algorithm picks an input table $T_i$ and constructs a set of dyadic gap boxes that include all the maximal dyadic gap boxes in them. We assume that all the attributes have a specified domain and they have been mapped to an integer in $[1,n]$. The algorithm initializes two sets $D$ and $D'$ to empty set. Then for all the tuples $x\in T_i$ and for all the dyadic boxes $b$ that contain $x$, the algorithm adds $b$ to $D$. This step can be performed by enumerating all the gap boxes whose elements in different dimensions are prefixes of the ones in $x$. 

Furthermore, for every box $b$ that is added to $D$ and every dimension $a$ of $b$ that does not span the whole space (is not $\lambda$), it creates a dyadic box by flipping the last bit of $a$ in $b$ and adds it to $D'$. For example, if $b = (10,\lambda,111)$, then the algorithm adds the dyadic boxes $(11,\lambda,111)$ and $(10,\lambda,110)$ to $D'$. At the end, the algorithm returns $D'\setminus D$. Algorithm \ref{alg:preprocessing} shows a pseudo-code of the preprocessing step.  The following algorithm will output at most $d (\log(m)+1)^d$ dyadic gap boxes which include all the maximal dyadic gap boxes. 

\begin{algorithm}[H]
    \caption{Algorithm for dyadic gap box construction}
    \label{alg:preprocessing}
    \begin{algorithmic}[1]
    \Procedure{Construct Gap Boxes}{$T_i$}
        \State \textbf{Input:} An input table $T_i$
        \State \textbf{Output:} A set of dyadic gap boxes including all maximal dyadic gap boxes.
        \State $D \gets \emptyset$
        \State $D' \gets \emptyset$
        \For{$x \in T_i$}
            \For{all dyadic boxes $b$ such that $x \in b$}
                \State $D \gets D \cup \{b\}$
                \For{all attributes $a$ such that $b_a \neq \lambda$}
                    \State $b' \gets b$
                    \State Flip the last bit of $b'_a$
                    \State $D' \gets D' \cup \{b'\}$
                \EndFor
            \EndFor
        \EndFor
        \State Return $D' \setminus D$.
    \EndProcedure
    \end{algorithmic}
\end{algorithm}

\begin{theorem}
\label{thm:number_of_box}
Let $B^i_d$ be the output of Algorithm \ref{alg:preprocessing} for table $T_i$ and let $B_d=\bigcup_i B^i_d$. Furthermore, let $B^i$ be the set of all the gap boxes for table $T_i$ and $B=\bigcup_i B^i$. Then,
\begin{enumerate}
    \item $|C(B_d)| = O(2^d \log^d(n)) |C(B)|$ 
    \item $|B_d| = O(d t \log^{d}(m))$ 
\end{enumerate}
\end{theorem}

Note that in the above theorem, $C(B)$ is the smallest possible certificate over any set of gap boxes. Therefore, we can conclude that Algorithm \ref{alg:preprocessing} produces a set of dyadic gap boxes that is efficient both in term of the size of the index and the size of the certificate.





\subsection{Dyadic Tree Data Structure}
    \label{sec:ds}
    
For our algorithm we need a data structure to efficiently store a set of dyadic boxes and search if a given dyadic box is included in the set or not. To store the dyadic boxes, first we consider each dyadic boxes as a string from the alphabet $\{0,1,\lambda, ,\}$ (the last element of the alphabet is a comma.) 
For example, a dyadic box of form $(00,\lambda,1)$ can be represented as the string ``$00,\lambda,1$". Then we use a Trie (prefix tree) to store and search the strings representing the dyadic boxes. Since there is a one to one relation between the dyadic boxes and their string representations, this approach will be able to correctly store the dyadic boxes and search for any query box.

Note that the length of every string representing a dyadic box is at most $O(d\log(n))$.
Thus the time needed for constructing the trie in the most straight forward manner is  $O(kd\log(n))$ where $k$ is the number of dyadic boxes, and the time to search for a dyadic box is $O(d \log(n))$.

\subsection{Chan's Algorithm for Klee's Measure Problem }
In this section we explain the general outline of the Chan's algorithm~\cite{chan2013klee}. Given a set of $n$ boxes in $d$ dimensional space, Chan's algorithm returns the volume of the union of the boxes in time $O(n^{d/2})$. We later use this algorithm to sample an uncovered point uniformly at random.
You may find the general description of the algorithm in Algorithm \ref{alg:chans}. The input of the algorithm is a set of boxes $B$ and a boundary box $S$ that represents the whole space. The output is the volume of the union of $B$ intersecting with $S$.

\begin{algorithm}[H]
    \caption{Chan's Algorithm}
    \label{alg:chans}
    \begin{algorithmic}[1]
    \Procedure{measure}{$B, S$}
        \State If $|B|$ is bellow a constant return the answer directly.
        \State Simplify $B$
        \label{step:chans:simplify}
        \State Cut $S$ into two sub-cells $S_1$ and $S_2$.
        \label{step:chans:cut}
        \State $B_1 \gets $ boxes in $B$ that have intersection with $S_1$
        \State $B_2 \gets $ boxes in $B$ that have intersection with $S_2$
        \State Return $\text{measure}(B_1,S_1) + \text{measure}(B_2,S_2)$
    \EndProcedure
    \end{algorithmic}
\end{algorithm}

The reason that the algorithm gets a good time complexity lies in steps \ref{step:chans:simplify} and \ref{step:chans:cut}. The simplification step $B$ removes any box that is equivalent to slabs spanning in all dimensions except for one when restricted to $S$. When removing these boxes, the algorithm changes $S$ and all the boxes in $B$ that has intersection with the slabs and removes the intersection from them. Note that this step doesn't increase the number of boxes in $B$.

After removing the slabs and simplifying the boxes, the algorithm cuts $S$ into two disjunctive sub-cells $S_1$ and $S_2$ and recursively calls the same function on those subproblems. The cut is done in a way that balances the number of faces of the boxes in $B$ that are present in $S_1$ and $S_2$ and it gives different weight to each of them based on the dimensions they are spanning.

The main properties of Chan's algorithm that are useful for us in order to uniformly sample a point are the following:
\begin{enumerate}
    \item $S_1$ and $S_2$ are disjunctive.
    \item There is a one to one relation between the uncovered points in $S_1$ and $S_2$ and the uncovered points in $S$.
\end{enumerate}

Using the above properties, it is very easy to sample $k$ uncovered points uniformly at random. We initialize $S$ to be the cross product of the domains of the attributes (having all possible tuples). Then we run Chan's algorithm to measure the number of uncovered points in $S$. The algorithm generates a list $L$ containing $k$ random positive integers smaller than the number of of uncovered points and runs the Chan's algorithm again; however, this time the input to the algorithm also has $L$ in a sorted array and the volume of uncovered points in the previous recursions (initially $0$). Then the leafs of the recursive call returns an uncovered point uniformly at random for any of the numbers in $L$ that is between the number of uncovered before this leaf and after this leaf. Algorithm \ref{alg:uniform} describes the process. 

\begin{algorithm}[H]
    \caption{Uniform Sampling Algorithm}
    \label{alg:uniform}
    \begin{algorithmic}[1]
    \Procedure{Sample}{$B, S, L, V$}
    
        \If{$|B|$ is bellow a constant}
            \State $V' \gets$ volume of uncovered points in $S$.
            \State $k' \gets |\{x | x \in L \text{ and } V<x\leq V + V'\}|$
            \label{step:binary:search}
            \State $P \gets$ $k'$ uniformly sampled uncovered points in $S$.
            \State return $(P, V')$
        \EndIf
        \State Simplify $B$ and $S$
        \label{step:uniform:simplify}
        \State Cut $S$ into two sub-cells $S_1$ and $S_2$.
        \State $B_1 \gets $ boxes in $B$ that have intersection with $S_1$
        \State $B_2 \gets $ boxes in $B$ that have intersection with $S_2$
        \State $(V_1, P_1) \gets$ Sample$(B_1, S_1, L, V)$
        \State $(V_2, P_2) \gets$ Sample$(B_1, S_1, L, V+V_1)$
        \State Map coordinates of $P_1$ and $P_2$ according to $S$ before step \ref{step:uniform:simplify}. 
        \label{step:uniform:mapping}
        \State Return $(V_1 + V_2,P_1 \cup P_2)$
    \EndProcedure
    \end{algorithmic}
\end{algorithm}

Note that since we simplify and cut $S$, the coordinates of the points returned by the recursive call are not the same as their original coordinate. Therefore, we need to perform step $\ref{step:uniform:mapping}$ to change their coordinates.

\begin{theorem}
\label{thm:uniform_sampling}
The uniform sampling algorithm can return $k$ uniformly at random selected uncovered points in time $O(\log(k)|B|^{d/2} + kd^2\log^2(|B|))$.
\end{theorem}
\begin{proof}
The only extra operations that Algorithm \ref{alg:uniform} is performing compared to Chan's Algorithm is the sampling of the points and mapping the coordinates of the sampled points in the recursion. Step \ref{step:binary:search} is done in all of the leafs and it takes $O(\log(k))$. The sampling step happens once per sampled point and it takes $O(d)$. The only remaining step to analyse is the mapping step. Mapping can be done in $O(d\log(|B|))$ for each sampled point, because all we need to know for each dimension is the location of the slabs that are removed. The number of times each sampled point goes through the mapping phase is the depth of the recursion tree. Analysis of Chan's algorithm shows that the depth of the recursion is $O(\log(|B|)d)$. Therefore, the total time that step \ref{step:uniform:mapping} takes in all of the recursions is $O(k \log^2(|B|) d^2)$.
\end{proof}

\section{The Welltris Algorithm}
Given a join query and the dyadic gap boxes of all relations, 
we propose the following algorithm for estimating the row count of a join in time proportional to the certificate size of that join. The proposed algorithm finds a $(1+\epsilon)$ approximation of the join size with probability at least $1-\delta$. Consider an input table $T_i$ and a gap box $b$ of $T_i$. We say $b$ covers a point $p$ if the projection of $p$ onto the subspace, with the dimensions in  $T_i$, lies inside $b$.

\paragraph*{Algorithm Welltris Description}
\begin{enumerate}
    \item Initialize the set of selected gap boxes $E$ to be empty and $B_d$ to be set of dyadic boxes of all the relations.
    \item Uniformly select $k=\frac{4}{\epsilon}(\log(|B_d|)+\log(\frac{1}{\delta}))$ random points with replacement independently that are not covered by the boxes in $E$.
    \label{step:sampling}
    \item Check the points one by one to find the dyadic gap boxes in $B_d$ that covers them. 
    \label{step:checking_boxes}
    \item For one of the points that are covered by boxes in $B_d$, add all the dyadic gap boxes in $B_d$ that covers it to $E$.
    
    \label{step:adding_new_boxes}
    \item If none of the points was covered by a dyadic gap box in $B_d$, return the volume of the points not covered by $E$ as the estimate of the join size.
\end{enumerate}

In the following theorem statement recall that $B$ is the set of all dyadic gap boxes of all tables; see Theorem~\ref{thm:number_of_box}.

\begin{theorem} 
Let $C(B)$ be the smallest certificate over all choices of gap boxes and $C=|C(B)|$, then given the dyadic gap boxes constructed in the preprocessing algorithm (Algorithm~\ref{alg:preprocessing}), Algorithm Welltris can return a $1+\epsilon$ estimation of the join size with probability at least $1-\delta$ in time $O(\frac{1}{\epsilon} \log(1/\delta) \cdot \log(m)  \cdot 2^{d^2/2+d} \cdot \log^{d^2+d}(n) \cdot C^{d/2+1})$.
\end{theorem}
\begin{proof}
First, we prove the approximation guarantee and then we prove the time complexity. The approximation guarantee can be proven by applying Chernoff bound and union bound.
Let $U_i$ be the set of uncovered points at iteration $i$ and let $|J|$ be the actual join. 
For an arbitrary iteration $i$, using the fact that $k$ points are sampled independently, we will prove that if $|U_i| \geq (1+\epsilon)|J|$ then the probability that our algorithm stops in that iteration and returns $|U_i|$, is at most $\frac{\delta}{|C|}$. Then using union bound over all possible iterations, whose number is bounded by $|C|$, we will show that the probability that our algorithm stops at any iteration for which $|U_i| > (1+\epsilon)|J|$ is at most $\delta$.

Fix an iteration $i$. Let $p_1, p_2, \dots, p_k$ be the points that the algorithm samples at iteration $i$, and let $X_1,X_2, \dots, X_k$ be a Bernoulli random variables such that $X_i$ is $1$ if and only if $p_i \notin J$ and let $X=\sum_{j=1}^k X_j$. Then we have $\Pr[X_j = 1] = \frac{|U_i \setminus J|}{|U_i|} \geq \frac{\epsilon }{1+ \epsilon} \geq \frac{\epsilon}{2}$ since $|U_i| > (1+\epsilon) |J|$. Note that the algorithm stops if $X = 0$. Using the fact that $\{X_j\}_{j \in [k]}$ are independent, we have
\begin{align*}
    \Pr[X=0] =  \prod_{j \in [k]} \Pr[X_j = 0] \leq ( 1 - \frac{\epsilon}{ 2})^k   \leq  \exp{(-\frac{k\epsilon}{2})}.
\end{align*}
Plugging in $k=\frac{4}{\epsilon}(\log(|B_d|)+\log(\frac{1}{\delta}))$ we have
$
\Pr[X=0] \leq \frac{\delta}{|B_d|}.
$

Note that the at each iteration of the algorithm, one of the boxes in the certificate of the dyadic boxes $C(B_d)$ is added to $E$, therefore the maximum number of iterations is $|C(B_d)|$ which is smaller than $|B_d|$. Therefore, using union bound over all the possible iterations, we can conclude that with probability $1-\delta$ the algorithm stops when $|U_i| \leq (1+\epsilon) |J|$ and it returns a $(1+\epsilon)$ approximation of the join size.

The time complexity of the algorithm depends on the time needed to sample the points, the time needed to find the boxes covering the sampled points, and the number of iterations. Let $C_d$ denote $|C(B_d)|$ and $C$ denote $|C(B)|$. Furthermore, let $E_i$ denote the number of boxes in $E$ at iteration $i$. At each iteration, based on Lemma \ref{lemma:dyadicbox_containing_dyadicbox}, at most $O(\log^d(n))$ boxes are added to $E$ including one of the boxes in $C_d$; therefore, using Theorem \ref{thm:number_of_box}, we can conclude that the number of iterations is at most $C_d$, and for all iterations we have $E_i=O(\log^d(n) C_d)$. 

Using Theorem \ref{thm:uniform_sampling}, the sampling of $k$ points in iteration $i$ takes $O(\log(k)E_i^{d/2} + k d^2 \log^2(E_i))$ which is at most $ O(k E_i^{d/2})$. Based on Theorem \ref{thm:number_of_box}, $|B_d| = d t m \log(m)$. Therefore, we can replace $k$ with the following upperbound: $$k=\frac{4}{\epsilon}(\log(|B_d|)+\log(\frac{1}{\delta})) = O(\frac{1}{\epsilon} \log(1/\delta) \log(m))$$ and $E_i$ with $O(\log^d(n) C_d)$ to derive the following time complexity for the sampling step in each iteration:
$$
    O(k E_i^{d/2}) \leq O(\frac{1}{\epsilon} \log(1/\delta) \log(m) \log^{d^2/2}(n) C_d^{d/2})
$$

Finding out if a point is covered by any of the dyadic boxes in $B_d$ can be done in time $O(d\log^{(d+1)}(n))$ using Dyadic Tree data structure by looking for all possible $O(\log^d(n))$ dyadic boxes that include the point; See Section~\ref{sec:ds}. Therefore, steps \ref{step:checking_boxes} and \ref{step:adding_new_boxes} can be performed in time $$O(k d\log^{d+1}(n)) = O(\frac{d}{\epsilon} \log(1/\delta) \log(m)  \log^{d+1}(n)).$$

Multiplying the time complexity of all three steps by the number of iterations gives us the total time complexity of 
$$O(\frac{1}{\epsilon} \log(1/\delta) \log(m) \log^{d/2}(n) C_d^{d/2+1} + \frac{d}{\epsilon} \log(1/\delta) \log(m)  \log^{d+1}(n) C_d),$$
and by replacing $C_d$ with $2^d \log^d(n) C$ we have
\begin{align*}
&O(\frac{1}{\epsilon} \log(1/\delta) 2^{d^2/2+d} \log^{d^2+d}(n) \log(m) C^{d/2+1} + \frac{d}{\epsilon} \log(1/\delta) \log(m) 2^{d} \log^{2d+1}(n) C)
    \\
    = &O(\frac{1}{\epsilon} \log(1/\delta) \log(m) 2^{d^2+d}\log^{d^2/2+d}(n) C^{d/2+1}).
\end{align*}
\end{proof}

We now explain how to modify Welltris to return $q$ points sampled uniformly
at random from the joined table $J$. 
The number of points $k$ sampled in each iteration is set to be $q$. 
Then in step \ref{step:checking_boxes}, every time that any of the the sampled point is not in any gap box, that point is returned  as one of the $q$ sampled points.
By rejection sampling, we can conclude these points are sampled uniformly at random from $J$.

\begin{corollary}
This modification of Welltris samples  $q$ points uniformly at
random from a join in time $O\left(\log(q)2^{d^2/2+d} \log^{d^2+d}(n)C^{d/2+1} + q d^4 2^d \log^{2d+1}(n)C\right)$.
\end{corollary}
\begin{proof}
The time complexity of steps \ref{step:sampling},\ref{step:checking_boxes}, \ref{step:adding_new_boxes} in each iteration is $O(\log(q)E_i^{d/2} + q d^2 \log^2(E_i) + q d\log^{d+1}(n))$, note that $E_i = O(2^{d} \log^{2d}(n) C)$, and we can replace $\log(E_i)$ with $d\log(n)$ because the maximum size of the certificate is $n^d$. Multiplying by the maximum number of iterations $C_d = O(2^{d} \log^d(n) C)$ we derive the claimed time complexity.
\end{proof}

\bibliographystyle{plain}
\bibliography{main}

\begin{thebibliography}{10}

\bibitem{Welltris}
{\em Welltris wikipedia page}.

\bibitem{AfshaniBC17}
Peyman Afshani, J{\'{e}}r{\'{e}}my Barbay, and Timothy~M. Chan.
\newblock Instance-optimal geometric algorithms.
\newblock {\em J. {ACM}}, 64(1):3:1--3:38, 2017.

\bibitem{alway2019domain}
Kaleb Alway.
\newblock Domain ordering and box cover problems for beyond worst-case join
  processing.
\newblock Master's thesis, University of Waterloo, 2019.

\bibitem{abs-1909-12102}
Kaleb Alway, Eric Blais, and Semih Salihoglu.
\newblock Box covers and domain orderings for beyond worst-case join
  processing.
\newblock {\em CoRR}, abs/1909.12102, 2019.

\bibitem{CHAN2010243}
Timothy~M. Chan.
\newblock A (slightly) faster algorithm for {K}lee's measure problem.
\newblock {\em Computational Geometry}, 43(3):243 -- 250, 2010.

\bibitem{chan2013klee}
Timothy~M Chan.
\newblock Klee's measure problem made easy.
\newblock In {\em IEEE Symposium on Foundations of Computer Science}, pages
  410--419, 2013.

\bibitem{grohe2006structure}
Martin Grohe.
\newblock The structure of tractable constraint satisfaction problems.
\newblock In {\em International Symposium on Mathematical Foundations of
  Computer Science}, pages 58--72. Springer, 2006.

\bibitem{khamis2016joins}
Mahmoud~Abo Khamis, Hung~Q Ngo, Christopher R{\'e}, and Atri Rudra.
\newblock Joins via geometric resolutions: Worst case and beyond.
\newblock {\em ACM Transactions on Database Systems (TODS)}, 41(4):22, 2016.

\bibitem{marx2013tractable}
D{\'a}niel Marx.
\newblock Tractable hypergraph properties for constraint satisfaction and
  conjunctive queries.
\newblock {\em J. ACM}, 60(6):42, 2013.

\bibitem{Roughgarden19}
Tim Roughgarden.
\newblock Beyond worst-case analysis.
\newblock {\em Commun. {ACM}}, 62(3):88--96, 2019.

\bibitem{unconditionalCoreset}
Alireza Samadian, Kirk Pruhs, Benjamin Moseley, Sungjin Im, and Ryan~R. Curtin.
\newblock Unconditional coresets for regularized loss minimization.
\newblock In {\em International Conference on Artificial Intelligence and
  Statistics, {AISTATS}}, volume 108 of {\em Proceedings of Machine Learning
  Research}, pages 482--492. {PMLR}, 2020.

\bibitem{Ullman}
Jeffrey Ullman, Hector Garcia-Molina, and Jennifer Widom.
\newblock {\em Database Systems: The Complete Book}.
\newblock Prentice Hall PTR, 2001.

\end{thebibliography}

\end{document}